\documentclass[10pt,aps,prl,showpacs,twocolumn]{revtex4-2}
\usepackage{amsmath,amssymb,graphicx}
\usepackage{mathptmx}
\usepackage{color}

\usepackage{addfont}
\addfont{OT1}{rsfs10}{\rsfs}

\usepackage{xcolor}

\usepackage{textcomp}

\usepackage{lineno}
\usepackage{physics}

\usepackage[breaklinks]{hyperref}
\usepackage{inputenc}

\usepackage[scr=rsfso,cal=zapfc,frak=euler,bb=ams]{mathalfa}
\begin{document}

\title{Robust three-dimensional high-order solitons and breathers in driven dissipative systems: a Kerr cavity realization}


\author{Yifan Sun$^{1,\dag}$}
\author{Pedro Parra-Rivas$^{1,\dag}$}
\author{Carles Mili\'{a}n$^{2}$}
\author{Yaroslav V. Kartashov$^{3}$}
\author{Mario Ferraro$^{1}$}
\author{Fabio Mangini$^{1}$}
\author{Mario Zitelli$^{1}$}
\author{Raphael Jauberteau$^{1}$}
\author{Francesco R. Talenti$^{1}$}
\author{Stefan Wabnitz$^{1,4}$}

\affiliation{$^1$ Department of Information Engineering, Electronics and Telecommunications, Sapienza University of Rome, Via Eudossiana 18, 00184 Rome, Italy}
\affiliation{$^2$ Institut Universitari de Matem\`{a}tica Pura i Aplicada, Universitat Polit\`{e}cnica de Val\`{e}ncia, 46022 Val\`{e}ncia, Spain}
\affiliation{$^3$ Institute of Spectroscopy, Russian Academy of Sciences, 
Troitsk, Moscow, 108840, Russia}
\affiliation{$^4$ CNR-INO, Istituto Nazionale di Ottica, Via Campi Flegrei 34, 80078}
\affiliation{*Corresponding author: yifan.sun@uniroma1.it}
\affiliation{$\dag$ These authors have contributed equally.}

\begin{abstract} 

We present a general approach to excite robust dissipative three-dimensional and high-order solitons and breathers in passively driven nonlinear cavities. Our findings are illustrated in the paradigmatic example provided by an optical Kerr cavity with diffraction and anomalous dispersion, with the addition of an attractive three-dimensional parabolic potential. The potential breaks the translational symmetry along all directions, and impacts the system in a qualitatively unexpected manner: three-dimensional solitons, or light-bullets, are the only existing and stable states for a given set of parameters. This property is extremely rare, if not unknown, in passive nonlinear physical systems. As a result, the excitation of the cavity with any input field leads to the deterministic formation of a target soliton or breather, with a spatiotemporal profile that unambiguously corresponds to the given cavity and pumping conditions. In addition, the tuning of the potential width along the temporal direction results in the existence of a plethora of stable asymmetric solitons. Our results may provide a solid route towards the observation of dissipative light bullets and three-dimensional breathers. 

\end{abstract}

\maketitle
Solitons are self-sustained localized packets of light or matter waves, capable of propagating unchanged in nonlinear media, owing to the interplay between diffraction/dispersion and nonlinear processes. These particle-like objects have been extensively studied in different areas of physics, such as Bose-Einstein condensates, plasmas, hadron matter, gravitation, and optics \cite{malomed_spatiotemporal_2005,dauxois_physics_2006,kartashov_frontiers_2019,Malomed_multi}. To date, most experiments on solitons have been carried out in one-dimensional (1D) and two-dimensional (2D) settings \cite{kartashov_frontiers_2019}. Whereas stable, steadily propagating 3D solitons (in optics they are usually called light bullets \cite{silberberg_collapse_1990}), theoretically predicted  over the realm of nonlinear science, and in different optical settings, remain yet elusive. This long-standing challenge is caused by high-order perturbations, which eventually cause the decay of 3D solitons, or by wave collapse, which is typical for materials exhibiting the ubiquitous cubic nonlinearity \cite{kartashov_frontiers_2019,berge_wave_1998-1,kuznetsov_bifurcations_2011}. Remarkably, observations of transient 3D solitons have been reported in the context of optics \cite{minardi_three-dimensional_2010,Renninger2013,panagiotopoulos_super_2015}: however, these states eventually decay, because of the above mentioned effects. Various strategies to delay or arrest 3D soliton decay have been proposed, including dynamical regularisation of collapse \cite{bree_regularizing_2017}, use of saturable, nonlocal, and competing nonlinearities \cite{Akhmediev1993, edmundson_robust_1992, bang_collapse_2002, mihalache_three-dimensional_2006,desyatnikov_three-dimensional_2000,mihalache_stable_2002}, rapid longitudinal variations of material parameters \cite{Matuszewski2004}, use of optical lattices 
\cite{Aceves1994,aceves1995energy,morsch_dynamics_2006, christodoulides_discretizing_2003,kartashov_soliton_nodate}, both static and twisted ones \cite{milian_robust_2019}, and other methods of wave confinement \cite{raghavan_spatiotemporal_2000,shtyrina_coexistence_2018}.

Amongst all physical systems, driven dissipative cavities offer the unique possibility to integrate higher order (otherwise detrimental) effects, such as higher-order dispersion and Raman scattering, into the soliton states, since these can be locked by the interplay between parametric losses and the driving source, thus preventing instabilities from their development \cite{parra2014third,milian2015solitons,milian2018clusters}. 
On the other hand, standard dissipative systems, like those described by the cubic-quintic Ginzburg-Landau or by passive-driven nonlinear equations, present multiple coexisting attractors, and in particular the one corresponding to a homogeneous flat (or basal) state \cite{fauve_solitary_1990,coullet_localized_2002}. In the presence of the homogeneous state, the excitation of one of the solitonic attractors is a nontrivial task, and it highly depends on the way the system is perturbed (see, e.g., discussion in Ref.\cite{guo2017universal}). Below, we refer to these localized solitonic attractors as dissipative solitons.

In this Letter, we unveil a rare paradigm for 3D dissipative soliton excitation, where the soliton itself is the only possible attractor of the system, even when it represents excited, higher-order state \cite{grelu_dissipative_2012}. This can be achieved by breaking the translational symmetries of the dissipative system -- a coherently driven Kerr resonator -- by to the introduction of a 3D parabolic confining potential. In the presence of the external uniform pump, this results in the emergence of different types of spatiotemporal dissipative solitons (STDSs), which represent nonlinear deformations of the various eigenmodes of the potential, including 'excited' ones, containing several nodes in their wavefunction. Among them, we predict the formation of high-order stable states, containing considerable contributions from several modes as determined by pump detuning from the cavity resonance. These states can be stable over a broad range of pump frequencies. The existence of such higher-order states in stable form is a remarkable and unexpected fact, since it is well-known that higher-order 3D solitons with complex spatiotemporal structures, such as vortex bullets \cite{Eilenberger2014,Leblond2007,Desyatnikov2005,MALOMED2019108},
and especially states with nodes, are fragile objects in both conservative and dissipative systems \cite{mihalache_stable_2006-1,Skarka2010,Veretenov2017,javaloyes2016cavity}
As such, their stabilization usually requires the presence of competing/nonlocal nonlinearities or dissipation mechanisms. We characterise the rich bifurcation structure of STDSs, as well as their stability, in different regimes of operation. We also show that the nonlinear cavity system also supports robust breather solutions, which emerge from Hopf bifurcations and oscillatory instabilities of the STDSs. From the application side, our approach may lead to the generation of robust optical frequency combs \cite{PASQUAZI20181, pasquazi:22} based on 3D dissipative solitons in multimode nonlinear cavities, generalising recent results \cite{skry2018two,ivars2021reversible}.

The master equation describing the time evolution of light in driven, passive, diffractive and dispersive Kerr cavities with a 3D potential reads
\begin{equation} 
	\partial_t A= \mathrm{i}\nabla^2 A - \mathrm{i}\left(x^2+y^2+C\tau^2\right)A+ \mathrm{i}|A|^2A  - (\alpha+\mathrm{i}\delta)A+P,
\label{eq1}
\end{equation}
where $A(x,y,\tau,t)$ is the slowly varying amplitude of the electric field, $t$ is a "slow time", $\nabla^2=\nabla_{\perp}^2+\partial_\tau^2$, $\nabla_{\perp}^2=\partial_x^2+\partial_y^2$ accounts for diffraction, $\partial_\tau^2$ stands for group velocity dispersion, $\alpha$ is the loss coefficient (further we set $\alpha=1$), $\delta$ is the detuning between the driving laser and the closest cavity resonance, $P$ is the pump amplitude, and $x^2+y^2+C\tau^2$ is the parabolic potential, where $C$ controls temporal confinement. The transverse potential $(x^2+y^2)$ is associated with a parabolic graded-index profile, while the temporal part $C\tau^2$, crucial for our findings, can be introduced by intracavity synchronous phase modulation \cite{Mecozzi:92,Tusnin2020,Englebert2021}.
Parabolic potentials have been previously considered in conservative systems for studying, e.g., vortex solitons in Bose-Einstein condensates \cite{malomed_stability_2007,lashkin_stable_2008}, and in dissipative systems for studying mode-locked nanolasers\cite{Sun2019,Sun2020}, multimode fiber lasers \cite{Kalashnikov_2022}, and stabilising 1D solitons \cite{sun2022}.

In the context of driven resonators, Eq.(1) is a spatiotemporal generalization of the well-known Lugiato-Lefever equation \cite{Lugiato1987}, or the temporal Haelterman-Trillo-Wabnitz equation \cite{haelterman_dissipative_1992}, with a potential. 
This equation is one of the simplest models describing the evolution of a complex field in the presence of dissipation and driving, and it has been used within different physical systems, e.g. in condensed matter and plasma physics \cite{morales_ponderomotive-force_1974,kaup_theory_1978}. 
Below, we thoroughly analyze the stationary solutions of Eq.(1) and their stability \footnote{Stationary solutions were computed with path-continuation methods, by using the open distribution software AUTO-07p \cite{Doedel2009}, as well as the Newton-Raphson method. The stability of the STDSs was explicitly computed by solving the linear stability eigenvalue problem derived from Eq.~(\ref{eq1}), and further confirmed via extensive time propagation simulations.}.


\begin{figure}
\centering
\includegraphics[scale=1]{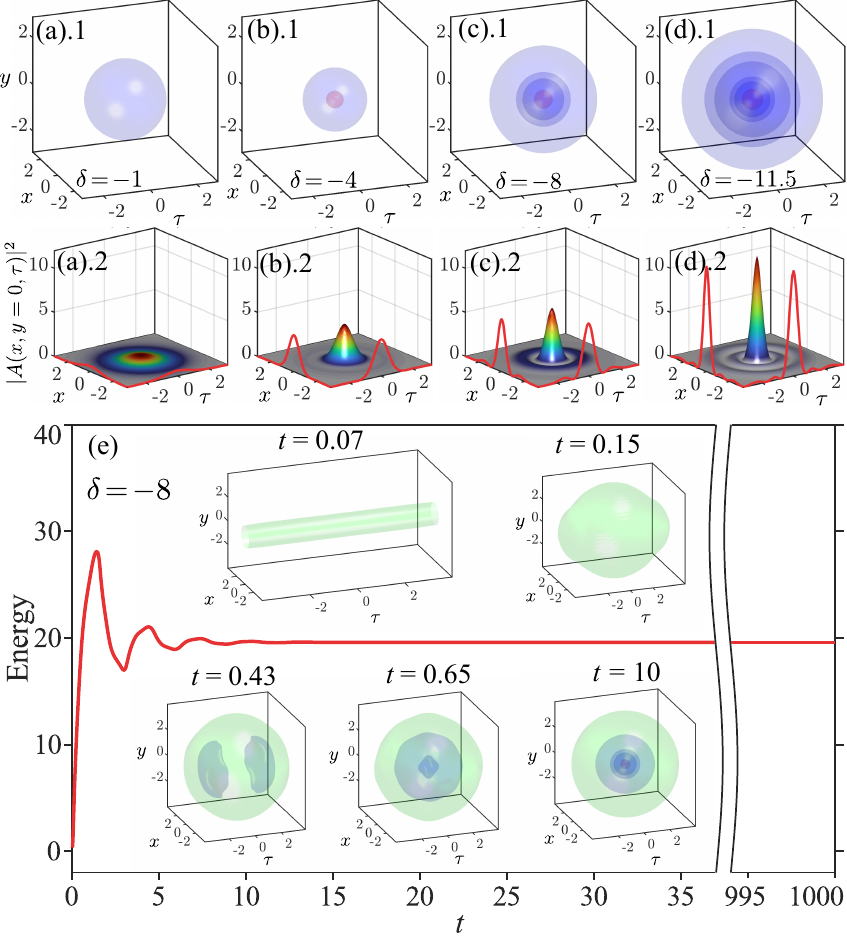}
\caption{(a-d) STDS solutions with $P=0.75$, $C=1$ and $\delta=-1,\, -4\,-8,\,-11.5$. Top: 3D representations; bottom: wave-functions cross sections, $|A(x,y=0,\tau)|^2$ at $y=0$. (e) Temporal evolution of the field energy, associated with the dynamical generation of the light bullet in (c), when exciting the cavity with a weak CW state (see text). All 3D plots in this figure consist of constant intensity surfaces at the values 3 (red), $0.1$ (blue), and $0.01$ (green). Blue surfaces mark approximate locations of the bullet's nodes, illustrating their higher-order features, as in (c)-(d).}
\label{fig1}%
\end{figure}	

Equation~(\ref{eq1}) with $C=1$ supports a variety of spherically symmetric STDSs: examples are illustrated in Fig.~\ref{fig1}(a)-(d) for $P=0.75$: by varying $\delta$ one obtains STDSs with different number of radial nodes, resulting in different intensity rings, as shown in Figs. 1(a-d).2. Our striking finding is that any of these stable states can be deterministically excited from an arbitrary input condition because they represent the unique stable attractor of the system for a given $\delta$. This fact is illustrated by Fig.1(e), which shows the evolution of the total intracavity energy, $E(t)\equiv\iiint |A(x,y,\tau,t)|^2\dd x\dd y\dd \tau$, starting from the weak CW input seed $A(x,y,\tau,0)=0.1{\rm exp}\left(-(x^2+y^2)/w^2\right)$ with $w^2=2/\ln 2$ (exactly the same final state is reached with any other arbitrary input). The initial stages of the field evolution feature a spatiotemporal confinement (see insets for $t=0.07$ and $t=0.15$, and visualization I), followed by the energy growth, which, despite the strong initial asymmetry (cf. inset at $t=0.43$), eventually converges, for $t\gtrsim10$, to the stable STDS presented in Fig.~\ref{fig1}(c). This state remains stable for extremely long propagation simulations ($t>1000$), in agreement with the predictions of the linear stability analysis (see below).


\begin{figure}
\centering
\includegraphics[scale=1]{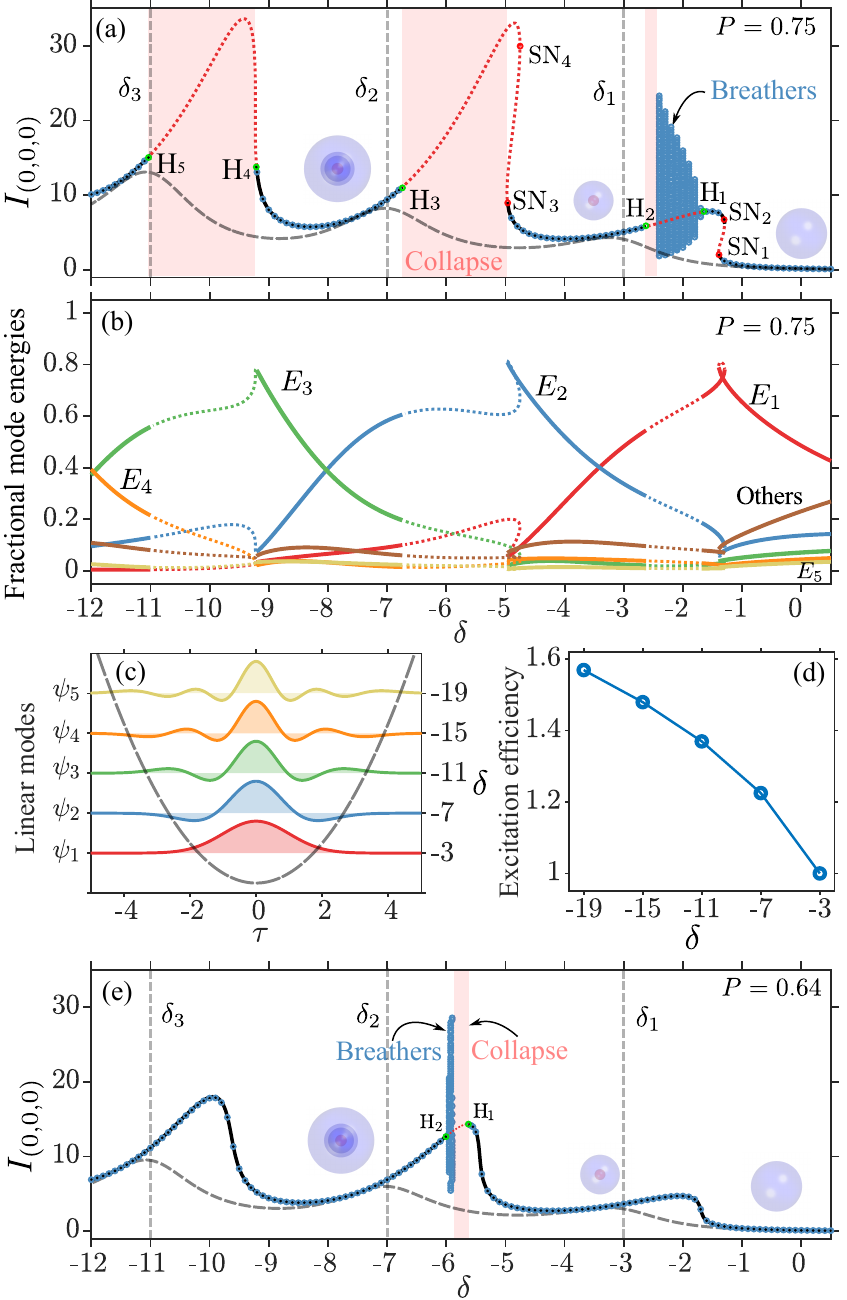}
\caption{(a) Bifurcation diagram showing the central peak intensity, $|A(x=0,y=0,\tau=0)|^2$ vs. detuning for $P=0.75$ and $C=1$. Solid black (dashed red) lines, obtained by numerical continuation, represent stable (unstable) STDSs. Blue circles superimposed on solid lines are obtained by propagation simulations. The gray dashed curve corresponds to the linear states of the cavity; vertical dashed lines at $\delta_1=-3$ ,$\delta_2=-7$, $\delta_3=-11$ mark the position of the resonances [as predicted from Eq.(2)]. Points $\textrm{H}_m$ mark the Hopf bifurcation thresholds, red areas mark the collapse regions, and the sparse blue circles mark the breathers minimum/maximum amplitudes. (b) Normalized modal energies of light bullets in (a), obtained by decomposition into the basis formed by the linear modes, $\psi_n$, of the potential. (c) The first five modes $\psi_n$ and their eigenvalues, $\delta_n$. (d) Mode excitation efficiency vs. detuning. (e) Bifurcation diagram analogous to (a) for $P=0.64$.} 
\label{fig2}
\end{figure}

Physical insight into the Kerr cavity system is presented below, by analysing the nonlinear solutions of Eq.(1), alongside with their stability. Figure~\ref{fig2}(a) shows the bifurcation structure associated with the spherically symmetric ($C=1$) STDSs for $P=0.75$, where the light bullet's peak intensity is plotted as a function of detuning [red/black curves]. Bullets display a multi-resonant behavior for $\delta<0$, which is inherited from the linear system, as revealed by the solutions of Eq.(1) with the nonlinear term omitted, shown by the grey curves. Linear resonances (local maxima) occur at values of cavity detuning that correspond to the eigenvalues, $\delta_n$ ($n=1,2,3...$), associated to the modes of the potential, $\psi_n$, which obey  
\begin{equation}
\delta_n \psi_n=[\nabla^2-(x^2+y^2+C\tau^2)]\psi_n
\label{eq_linear}.
\end{equation}
The first five modes are shown in Fig 1(c), together with their eigenvalues (right axis): as can be seen, $\psi_n$ presents $n-1$ radial nodes (below we use the normalisation $\iiint |\psi_{n}(x,y,\tau)|^2 \dd x\dd y\dd \tau=1$). Hence, solutions with a different number of nodes (different $n$) are associated with the presence of multiple resonances.
It is also apparent in Fig. 2(a) that consecutive linear resonances which result for decreasing values of $\delta$ have a progressively larger amplitude. We may relate their amplitudes to a modal excitation efficiency under the driving $P$, which can be evaluated from the integral $M_{n} = \iiint P\psi_{n}(x,y,\tau) \dd x\dd y\dd \tau$. The normalized value $M_{\mathrm{nor},n}=M_{n}/M_{1}$ is plotted in Fig.~\ref{fig2}(d) for the first five modes, i.e., from $\psi_1$ to $\psi_5$. The predicted increase of $M_{\mathrm{nor},n}$ with $n$ qualitatively explains the behavior of both linear and nonlinear resonances (which are associated to the amplitude of the 3D light bullets) in Fig.2(a), exhibiting higher intensities as the cavity detuning decreases.

Importantly, the 3D bullets found here have a rich multimode nature, which can be unveiled by expanding the nonlinear solutions on the basis of the 3D linear modes $\psi_n$: $A_{STDS}=\sum_{n\geq1}C_n\psi_n$, where the expansion coefficients are $C_{n} = \iiint A(x,y,\tau)\cdot \psi_{n}(x,y,\tau) \dd x\dd y\dd \tau$. Figure~\ref{fig2}(b) depicts the energy decomposition of the STDSs from Fig.~\ref{fig2}(a) into the linear mode base, by showing the normalized modal energies $E_n=|C_n|^2/\sum|C_n|^2$ as a function of $\delta$. For example, the bullet shown in Fig.1(c) ($\delta=-8$) forms, as shown in Fig.~\ref{fig2}(b), due to the strong hybrid contribution of high-order modes $\psi_{2,3}$ ($\sim80\%$), plus other modes ($\sim20\%$). The bullet is stable, and its stability persists when $\delta$ is decreased, until entering into a region where spatiotemporal collapse occurs, i.e., in-between points H$_4$ and H$_5$ in Fig.~\ref{fig2}(a). For even lower values of the detuning, $\delta<\delta_{\rm{H}_5}$, higher-order stable STDSs such as that shown in Fig.~\ref{fig1}(d) arise, due to the locking of even higher dominant modes $\psi_{3,4}$: this is confirmed by the mode decomposition plot in Fig. 2(b). Multimodal richness increases with decreasing $\delta$, consistent with the fact that additional linear modes may be excited. Remarkably, even though the bullets' morphological complexity increases with a decrease of $\delta$, wide stability domains persist in the valleys between consecutive resonances. Such stability domains broaden whenever the driving amplitude is reduced, as shown in Fig. 2(e): this case is similar to Fig. 2(a), but with $P=0.64$. Here, the STDSs remain stable for almost all values of $\delta$.

The most exciting property of the STDS bifurcation diagrams [Figs. 2(a),(e)] is that light bullets do not coexist with any trivial states, such as stable homogeneous or quasi-homogeneous solutions, which are commonly emerging as basal states in dissipative systems. In our cavity, homogeneous states cannot exist because of the removal of the translation symmetry, which is achieved by the three-dimensional potential. As a result, bullets are excited from any initial input conditions [cf. Fig. 1(e)], constituting a rare paradigm for dissipative soliton formation. This is the central result of this Letter.

\begin{figure}
\centering
\includegraphics[scale=1]{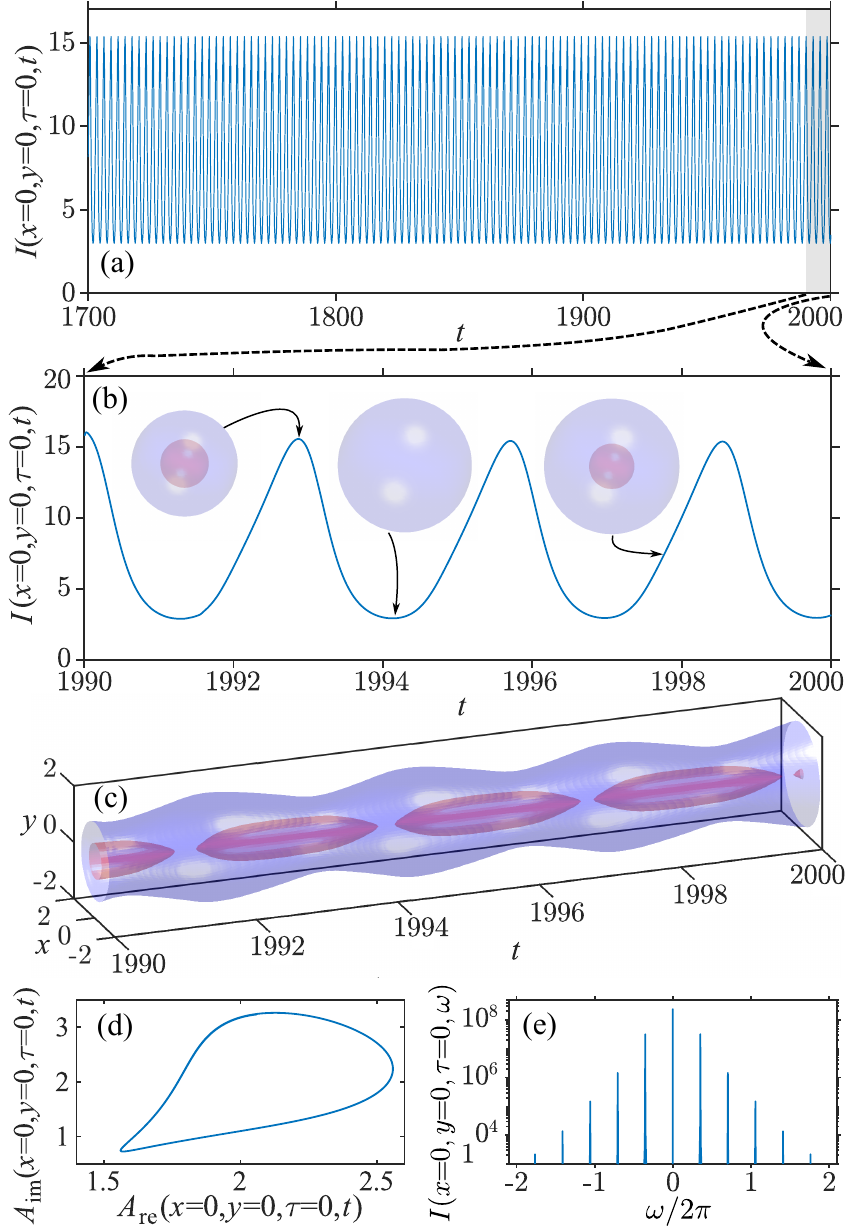}	
\caption{Time evolution of the breather with $P=0.75$, $\delta=-2$, $C=1$. (a) Peak intensity vs. time, shown only for the last $\sim106$ breathing periods obtained from a $t=2000$ long propagation simulation. (b) Zoom of (a) over the last few periods with instantaneous profiles shown by iso-surfaces at the levels $I=3$ (red) and $I=0.1$ (blue). (c) Time evolution of a cross section $I(x,y,\tau=0,t)$. (d) Closed orbits formed by the central real and imaginary parts of the complex field over 700 breathing periods [See visualization II]. (d) Fourier transform of the oscillations in (a), forming a comb of temporal frequencies.}
\label{fig3}
\end{figure}

While STDS between resonances tend to be stable, Fig. 2(a) shows that their stability may be lost for branches with high peak intensities. The stability thresholds correspond to saddle node bifurcation points, marked as $SN_m$ (m=1,2,3...), or to points $\textrm{H}_m$ (m=1,2,3...), where the solutions of Eq.(1) undergo supercritical Hopf bifurcations [e.g., H$_1$ at $\delta\approx-1.67$ and H$_5$ at $\delta\approx-11$ in Fig. 2(a)]: correspondingly, the STDS are subject to the oscillatory instability. Within the unstable regions, bullets may eventually collapse (red areas) or form stable 3D breathers, whose regular intensity oscillation ranges are indicated by the blue region in Fig.2(a).

Figure~\ref{fig3} shows the breather that is formed for $\{\delta,P,C\}=\{-2,0.75,1\}$. This stable dissipative state features 700 regular and perfectly periodic intensity oscillations and exactly periodic over the $t=2000$ long simulation shown in Fig. 3(a), for the last 300 time units. The close-up views of such evolution in Figs.~\ref{fig3}(b) and \ref{fig3}(c) illustrate the periodic modifications of the breather's 3D profile and cross section, respectively. Stable breathers correspond to multidimensional limit cycles, whose projection in the subspace $\{A_{\rm re}(x=0,y=0,\tau=0,t), A_{\rm in}(x=0,y=0,\tau=0,t)\}$ is illustrated in Fig.~\ref{fig3}(d) for a simulation extending up to $t=2000$. The Fourier transform of $I_{(0,0,0)}(t)$ [see Fig.~\ref{fig3}(e)] yields an equi-spaced comb of oscillation frequencies, with spacing $\Delta\omega\approx\omega/2\pi\approx0.3525$, leading to a breather period $T=\Delta\omega^{-1}\approx2.83$. Importantly, such breathers can also be deterministically excited by any arbitrary input field, illustrating once again the unique properties of this system for generating stable STDSs. 
The amplitude of breather oscillations increases when one tunes $\delta$ away from the Hopf bifurcation, where the breather has emerged [cf. Figs. 2(a) and 2(e)], until the breather starts collapsing (red areas in Fig. 2). This instability disappears when $\delta$ is tuned across the next Hopf bifurcation, e.g., H$_2$ at $\delta_{{\rm H}_2}\approx-2.6$ in Fig. 2(a). 

\begin{figure}
	\centering
	\includegraphics[scale=1]{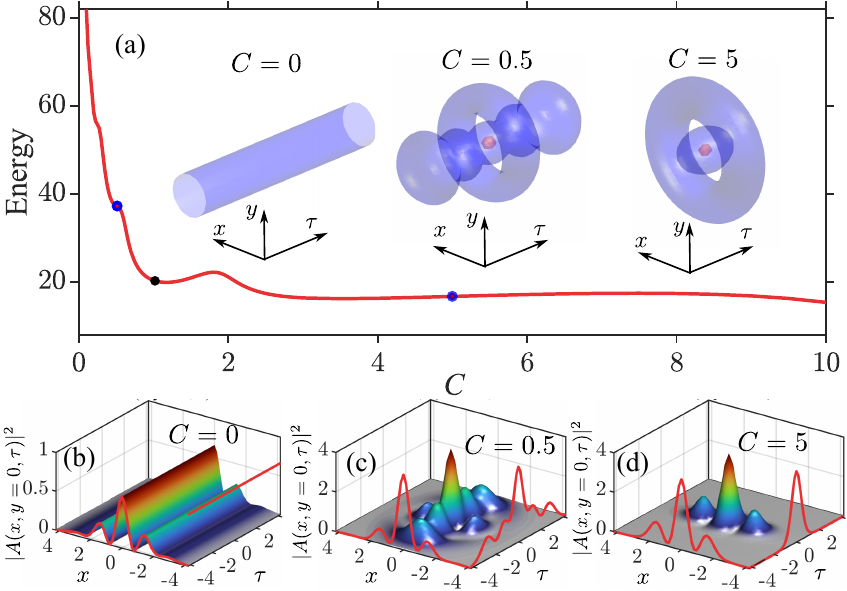}	
	\caption{
		(a) Energy of stable bullet solutions vs. $C$. (b-c) Intensity distribution $|A(x,y=0,\tau)|^2$ of three solutions (corresponding to three bullets in (a), where intensity shells are plotted: red: $I_1=3$, blue: $I_2=0.3$ with different $C$. The intensity distributions $|A(x=0,y=0,\tau)|^2$ and $|A(x,y=0,\tau=0)|^2$ are also plotted by red lines. Other parameters are $P=0.75$, $\delta=-8$. [see Visualization III]
  }	
	\label{fig4}%
\end{figure}

So far, we focused on STDSs appearing when the 3D potential is radially symmetric ($C=1$). The possibility of controlling the strength of the temporal confinement offers a powerful degree of freedom to generate a plethora of STDS with different shapes other than the spherically symmetric. Figure~\ref{fig4} shows the asymmetric bullets that are generated for $\delta=-8$ and $P=0.75$ at $C=0.5$ and $C=5$ [see insets in Fig. 4(a) and intensity cross sections at $y=0$ in Figs. 4(b)-(d)], together with the variation of the STDSs energy $E$ as a function of $C$ [Fig. 4(a)]. The bullet energy diverges at $C=0$, when it transforms into a uniform-in-$\tau$ cw state [left inset in Fig. 4(a) and Fig. \ref{fig4}(b)], which nevertheless preserves its complex multi-mode spatial structure, with several intensity oscillations. The increase of $C$ leads to a growing degree of temporal confinement and changes the morphology of the STDSs along the $\tau$-direction [see Visualization III]. The bullet energy $E$ typically decreases with $C$ [Fig. 4(a)]. At intermediate values of $C \sim 0.5$, one obtains STDSs with multiple maxima in the temporal dimension [Fig. \ref{fig4}(c)]. Further increase of $C$ results in strong temporal confinement for the STDSs [right inset in Fig.~\ref{fig4}(a) and Fig. \ref{fig4}(d)], while the bullet energy reaches an almost constant value around $E\approx17$. Remarkably, the asymmetric STDSs presented in Fig. \ref{fig4} are stable for all considered values of $C$.



In conclusion, we have introduced a novel paradigm for the deterministic excitation of robust STDSs in passive driven Kerr cavities with a 3D parabolic potential. The latter breaks the translational symmetries, and as a result STDS become the unique attractors of the system. Furthermore, stable high-order solitons and breathers, which are generally unstable, exist. We have characterized their bifurcation structure and stability, locating the thresholds between the stationary and breathing STDSs. Our general findings may stimulate further research on 3D solitons across several disciplines, including Bose-Einstein condensates, plasmas, and optics. In the latter, our work may pave the way for the long-sought experimental demonstration of truly stationary, long-living light bullets. In the context of soliton microcombs, the fact that STDS exist for negative (blue shifted) detuning should greatly facilitate their thermal locking \cite{Carmon2004}.

\begin{acknowledgements}
This work was supported by European Research Council (740355), Marie Sklodowska-Curie Actions (101064614, 101023717), Ministero dell'Istruzione, dell'Universita e della Ricerca (R18SPB8227). CM acknowledges support from the Spanish gouvernement via the Grant PID2021-124618NB-C21 funded by MCIN/AEI/ 10.13039/501100011033 and by “ERDF A way of making Europe”.
\end{acknowledgements}

\bibliography{references}

\end{document}